\documentclass[12pt,preprint]{aastex}
\usepackage{natbib}
\usepackage{lscape}

\newcommand{\ie}{\textit{i.e.}, }
\newcommand{\eg}{\textit{e.g.}, }
\newcommand{\etal}{\textit{et al.}}
\newcommand{\toss}{1996 TO$_{66}$}
\newcommand{\chisqr}{$\chi^{2}_{\nu}$}
\newcommand{\sigtot}{$\sigma_{total}$}
\newcommand{\sigstat}{$\sigma_{stat}$}
\newcommand{\sigsys}{$\sigma_{sys}$}

\shorttitle{Testing Accuracy and Precision of Existing \\Photometry Algorithms on Moving Targets}
\shortauthors{Sonnett \etal}

\begin{document}

\title{Testing Accuracy and Precision of Existing \\Photometry Algorithms on Moving Targets}

\author{S. Sonnett, K. Meech}
\affil{Institute for Astronomy, 2680 Woodlawn Drive, Honolulu, HI, 96822, USA; \\
University of Hawaii NASA Astrobiology Institute, Honolulu, HI, 96822, USA}

\author{R. Jedicke, S. Bus, J. Tonry, }
\affil{Institute for Astronomy, 2680 Woodlawn Drive, Honolulu, HI, 96822, USA}

\and

\author{O. Hainaut}
\affil{European Southern Observatory, Karl Schwarzschild Strasse, \\85748 Garching bei Munchen, Germany}

\begin{abstract}
Previous studies determining which astronomical photometry software is best suited for a particular dataset are usually focused on speed, source classification, and/or meeting a sensitivity requirement.  For faint objects in particular, the priority is given to maximizing signal-to-noise.  Photometry of moving targets offers additional challenges (i) to aperture photometry because background object contamination varies from image to image, and (ii) to routines that build a PSF model from point sources in the image because trailed field stars do not perfectly represent the PSF of the untrailed target.  Here, we present the results of testing several photometry algorithms (tphot, DAOPHOT, DoPHOT, APT, and multiple techniques within Source Extractor and IRAF's PHOT) on data for a faint, slow-moving solar system object with a known light curve.  We find that the newly-developed tphot software most accurately and precisely reproduces the object's true light curve, with particular advantages in centroiding, exclusion of contaminants from the target's flux, and fitting flux in the wings of the point-spread function.  
\end{abstract}

\keywords{Solar System: general --- Data Analysis and Techniques}

\clearpage

\section{Introduction \label{sec:intro}}

Photometry is the practice of measuring the flux and corresponding uncertainty of astronomical objects.  Many different photometry approaches have been developed, each of them optimized for a particular set of goals, and choosing which approach to adopt is a difficult task.  The science goals, intrinsic
nature of the target(s), and instrumental properties dictate which of the
dozens of available photometry algorithms might be most effective.  Some scientists, such as those with 
high-volume computing needs like the Sloan Digital Sky Survey group, develop their own codes, comparing them to popular algorithms 
\citep[\eg][]{2004AN....325..583I,2002SPIE.4836..350L,2000ASPC..203...50A}.  
Other groups try to optimize existing software for a specific task.  
For example, \cite{2007PASP..119.1462B} tested several algorithms for meeting 
LSST science requirements, and \cite{2000PASP..112..177F} assessed the 
performance of two algorithms in crowded fields.  

Photometry packages currently distributed for public use can be roughly divided 
into three groups: those that perform aperture 
photometry, those that construct a model point-spread function (PSF) from field 
objects, and those that blend the two.  It is now generally accepted that low signal-to-noise (S/N) object 
magnitudes are most accurately recovered through PSF-fitting since the errors are dominated by background uncertainty \citep[\eg][]{2003BaltA..12..243H,2007PASP..119.1462B}.
Conversely, high-S/N objects are best represented by aperture photometry, which are more forgiving than PSF-fitting techniques of out-of-focus 
frames and intrinsic morphological complexities (which may be time variable) in the PSF.  

In addition to accurately measuring flux, a second challenge to photometry is absolute calibration, which converts the flux measured in instrumental units (or instrumental magnitudes) into true apparent magnitudes.  Absolute calibration takes into account the zeropoint (the scaling factor in magnitude units), the atmospheric extinction, and the color conversion from those of the filters used to the standard color system.  We shall not discuss calibration of this dataset in this paper, which focuses only on the initial flux measurements.

Moving targets are typically comets and asteroids, which by virtue of their size, distance, and low reflectivity, are very faint; thus, observations of the primary science targets are typically low S/N, making optimal sky determination critical.  Unless imperfections in the optics distort the field, fixed point sources in the field should have exactly the same PSF that can most likely be described by a fitting function.  Moving targets, however, will produce a trailed PSF that may not be approximated by a function, with the amount of trailing dependent on the object's proper motion.  No publicly available photometry software was designed to handle trailed PSFs.  We conducted the first study that quantitatively determines which available photometry 
algorithm optimizes accuracy and precision for faint moving targets.  We familiarized ourselves with 15 different aperture or PSF-fitting algorithms and tested which one most accurately and precisely reproduced the true light curve of a moving object with a known rotational light curve.

\section{Observations \label{sec:obs}}

One way of testing a technique's accuracy is to use it on data with a known trend and measure
the root-mean-square (RMS) of the residuals against a model.  We acquired data for the 
Trans-Neptunian object \toss\, as part of a campaign to determine the object's light curve.  
As the purpose of this paper is simply to use this dataset to explore methodology, we 
will not comment on the campaign itself, nature of the target, or the details of the model and instead defer that information 
to an upcoming publication (Hainaut \etal, \textit{private comm.}).  

\begin{figure}[h]
\epsscale{0.65}
\begin{center}
\rotate
\plotone{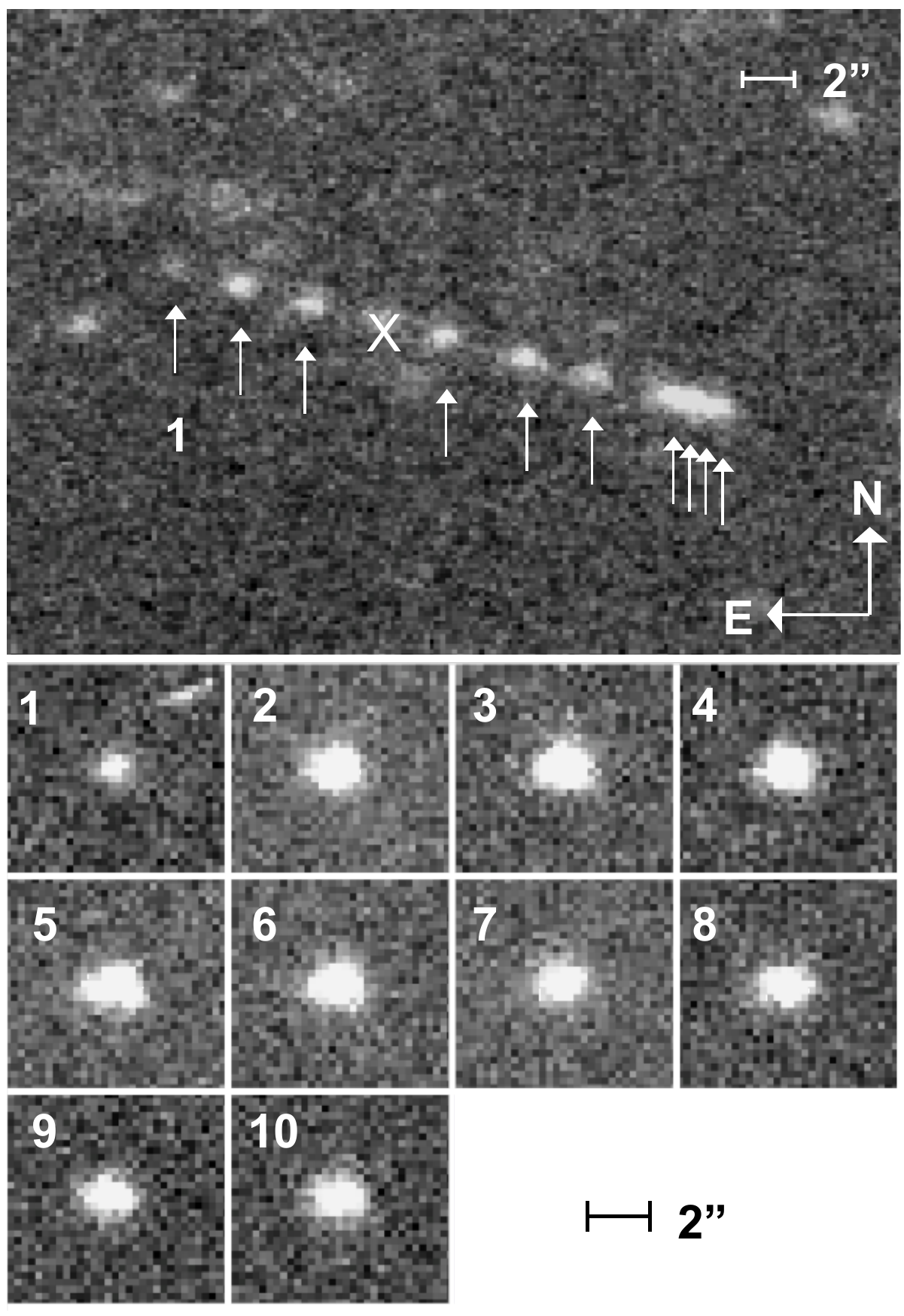}
\caption{Images used in analysis.  The top panel is the sky composite showing \toss's motion across the sky (indicated by arrows, with the first image indicated by a ``1''), showing the background in the target's vicinity.  The composite was made by shifting, mode-scaling, and summing all frames.  The data point indicated by an ``X'' was excluded because it was likely contaminated by a bad pixel.  The bottom panel shows postage stamps of the target in the individual frames used in analysis (ordered chronologically $1- 10$) in order to show the PSF morphology.  The stamps have the same brightness scale.  A cosmic ray hit $\sim 3.1\arcsec$ from the target's centroid in Frame 1 may have compromised some of the photometry, and minor focus problems manifested as slightly triangular PSFs are seen in Frames 5 and 6.  \label{fig:images}}
\end{center}
\end{figure}

The data were obtained on September 22, 2011 using the University of Hawai'i (UH) 2.2-m 
telescope with the Kron-Cousins $R$-band filter and the Tek ($2048 \times 2048$) CCD camera, which has a pixel scale of 0.219$\arcsec$/pixel.  The sky was not photometric due to occasional cirrus clouds.  We measured a median seeing of 
0.95$\arcsec$ FWHM, which varied by $\sim 20\%$ over the course of the \toss\, observations.  The focus was checked before beginning the \toss\, observations, but we could not change the focus later as the night progressed due to technical difficulties.  Consequently, a slightly triangular-shaped PSF indicative of poor focus was present in two of our images (Fig. \ref{fig:images}).
Exposure times varied between 900 and 1100 seconds (the median being $945s$) to 
achieve our S/N goal of $\sim 30$, which is still too faint to have well-defined PSF wings.  This was done in an attempt to achieve a constant S/N under changing transparency and seeing conditions.   

\begin{figure}[h]
\epsscale{0.85}
\begin{center}
\rotate
\plotone{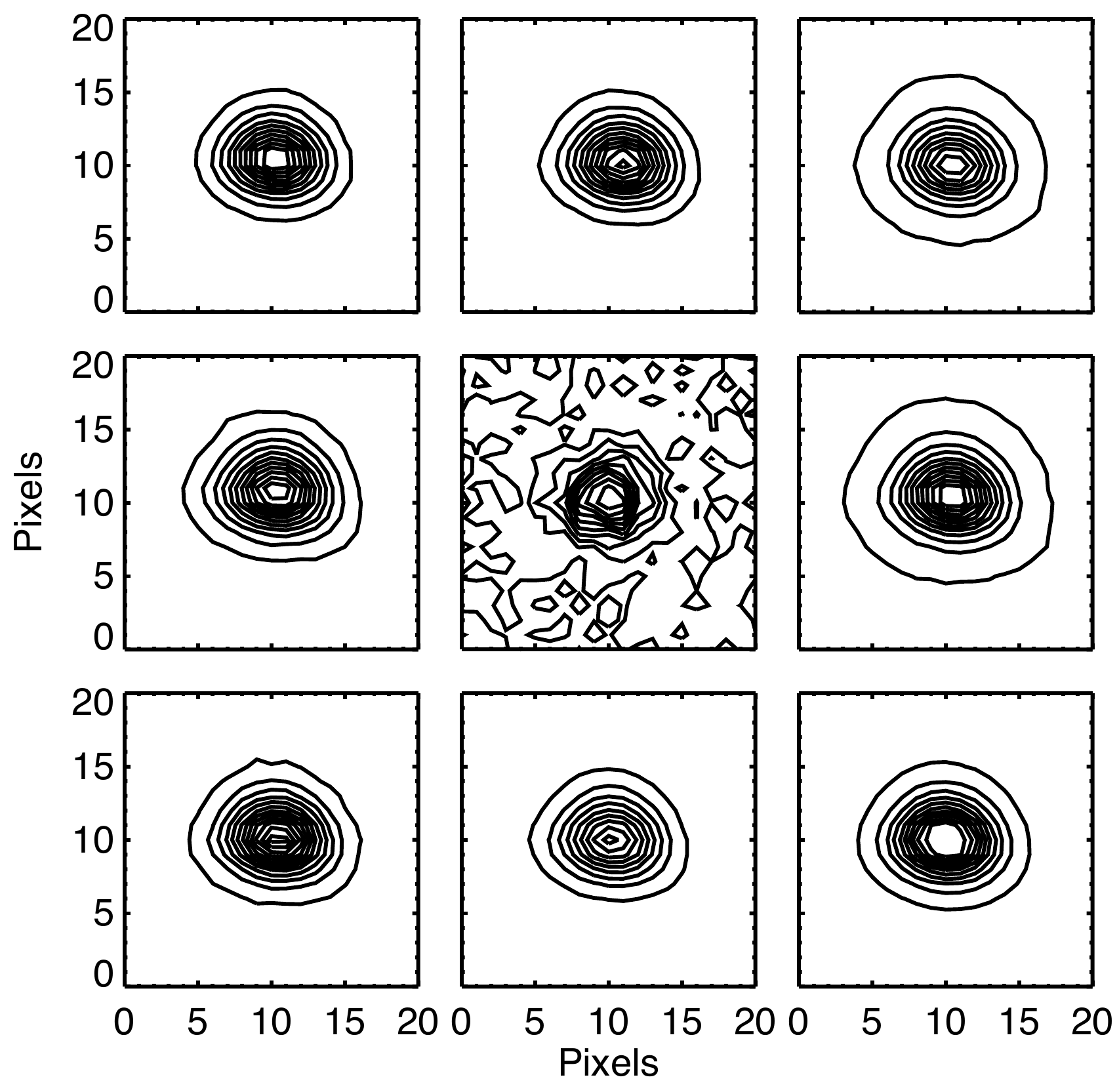}
\caption{Contour plots of the target (center panel) and surrounding bright field stars for Frame 5.  Based on tphot's best-fit PSF models, the trailing aspect is 1.5 for field stars and 1.1 for the target.  The PSF shapes are consistent across the image, indicating no spatial variation to a fraction of a percent.  \label{fig:contours}}
\end{center}
\end{figure}

Non-sidereal 
guiding was used so that the target was not trailed.  The target's motion averaged
-2.7$\arcsec/$hr in R.A. and -0.8$\arcsec/$hr in Dec., corresponding to 
$\sim 0.72\arcsec$ of trailing per frame for stars, or a theoretical PSF length-to-width ratio (or trailing aspect) of $\sim 1.8$ for field stars and 1.0 for the target.  The best-fit models using the tphot PSF-fitting algorithm, however, give trailing aspects of 1.5 for the field stars and 1.1 for the target; the difference between the theoretical and actual trailing aspect is likely attributable to telescope guiding problems (Fig. \ref{fig:contours}).  We stress that this amount of trailing is considered minimal compared to typical rates of inner solar system objects \citep{2012PASP..124.1197V}.  The medians of the peak and background counts were $\sim 3350$ and $\sim 2550$, respectively, and the median flux within a 7-pixel aperture was 23000 $\pm$ 600 counts.  We calculate a medan S/N of $\sim 35$ using the tphot algorithm (discussed in Section \ref{sec:tphotap}).  

Figure \ref{fig:images} shows both \toss's track along the sky and the target-centered sub-sections of the ten \toss\, images. The observations interleaved \toss\, with another target over 8.12 hours, until \toss\, reached airmass 2.0.  Preliminary results from Hainaut \etal$\;$(private comm.) give a light curve amplitude of $\sim 0.14$ magnitudes and a rotation 
period of $7.94\pm0.33$ hours.  The model light curve reproduces all data obtained as part of a multi-year, multi-telescope, multi-solar phase angle observational campaign to determine the light curve of \toss.  Our observations correspond to rotation phases ($\phi$) of $0.114-0.989$, or 87.5\% of the full rotational light curve.  One data point was consistently fainter than the model by $\gtrsim 0.1$ magnitudes regardless of the technique used, likely due to a bad pixel that was not identified during construction of a bad pixel mask, so we excluded it from analysis (Fig. \ref{fig:images}).  

\section{Initial Image Processing \label{sec:processing}}

We prepared the images for analysis using IRAF's ``CCDPROC'' package \citep{1986SPIE..627..733T}, applying bad pixel, bias, 
and overscan corrections and flat-fielding with dithered twilight sky frames.  To exclude field stars
from the bad pixel mask construction, we median-combined the brightest twilight flats to produce a 
stacked bright flat, then repeated that process for a stacked faint flat in the same filter.  We fed the ratio of the 
stacked bright to the stacked faint flat to the IRAF ``CCDMASK'' task, which computes a bad pixel mask 
from a ratio image, then used the ``fixpix'' task within CCDPROC to interpolate over bad pixels.  We noted that in two images, a bad pixel was relatively close to the target -- 8 pixels from the target centroid in Frame 3 (Fig. \ref{fig:images}), and 9 pixels from the target centroid in the Frame that was excluded from analysis (See Section \ref{sec:obs}).  The bad pixel regions were visually and quantitatively comparable to the local sky after correction (as it should be), so their proximity to the target was of no consequence to our photometry.  Afterward, we
fit a polynomial to the overscan for line by line bias subtraction and trimmed the data.  We then combined 20 bias frames taken at the beginning and end of the night to produce a stacked bias frame for subtraction.  

We acquired seven $R$-band twilight sky flats at the beginning of the night and six at the end, all below the limit of the linearity regime for the chip.  However, because the dust pattern changed multiple times within the 
\toss\, observation window, the twilight flats were not always representative of the field.  Consequently
in the flattened images, residuals from improperly-corrected dust donuts can be seen in a third of our
images, possibly influencing background determination for field stars in affected areas.  The residual dust 
donuts were not close enough to the target to affect the target photometry from any of the algorithms.  Lastly, 
we corrected for cosmic rays using the ``COSMICRAY'' task within IRAF's ``CRUTIL'' package.  
We found that a threshold of $3.8\sigma$ relative to noise in $R$-band images effectively removed most cosmic rays without falsely flagging real sources.  The entire reduction procedure from bad pixel to cosmic ray correction did not differ between photometry algorithms.  

After reduction, we measured the effective gain and read noise by doing a linear least squares fit to the square of the sky noise over the sky, all in units of ADU.  The slope of the fit is the inverse of the gain, and the y-intercept is the square of the read noise in ADU.  We measured a gain and read noise of $\sim1.31$ e$^{-}/$ADU and $\sim21$ e$^{-}$, respectively.  

\subsection{Detection, Differential Photometry, and Systematic Errors \label{sec:processing2}}

Field source detection for all photometry algorithms was done using Source Extractor (SExtractor), which identifies point sources as a user-defined minimum number of adjacent pixels (sharing either a border or a corner) that are above a specified detection threshold \citep{1996A&AS..117..393B}.  If the intensity distribution of a cluster of source pixels shows a distinct saddle point, the surrounding peaks are considered to belong to two separate objects.  SExtractor then measures the shape and centroid of each object, eliminates bright object artifacts, and further cleans the source list.  Because we find this rigorous detection method very effective at identifying, de-blending, and broadly classifying all field sources, we used it to produce a coordinate list for field stars.  

We restricted field stars for further analysis to those that had S/N $>\,$10 within an aperture diameter of $4\times$FWHM, as preliminarily determined by SExtractor's ``MAG\_BEST'' routine.  To correct for extinction and occasional clouds for each algorithm, we applied an offset correction for each frame to bring the field stars brightnesses to the same level.  This correction was computed by removing the median offset in bright field star magnitudes between frames.  

\begin{figure}[h]
\begin{center}
%\plotone{syserr.pdf}
\includegraphics[angle=0,width=6.5in]{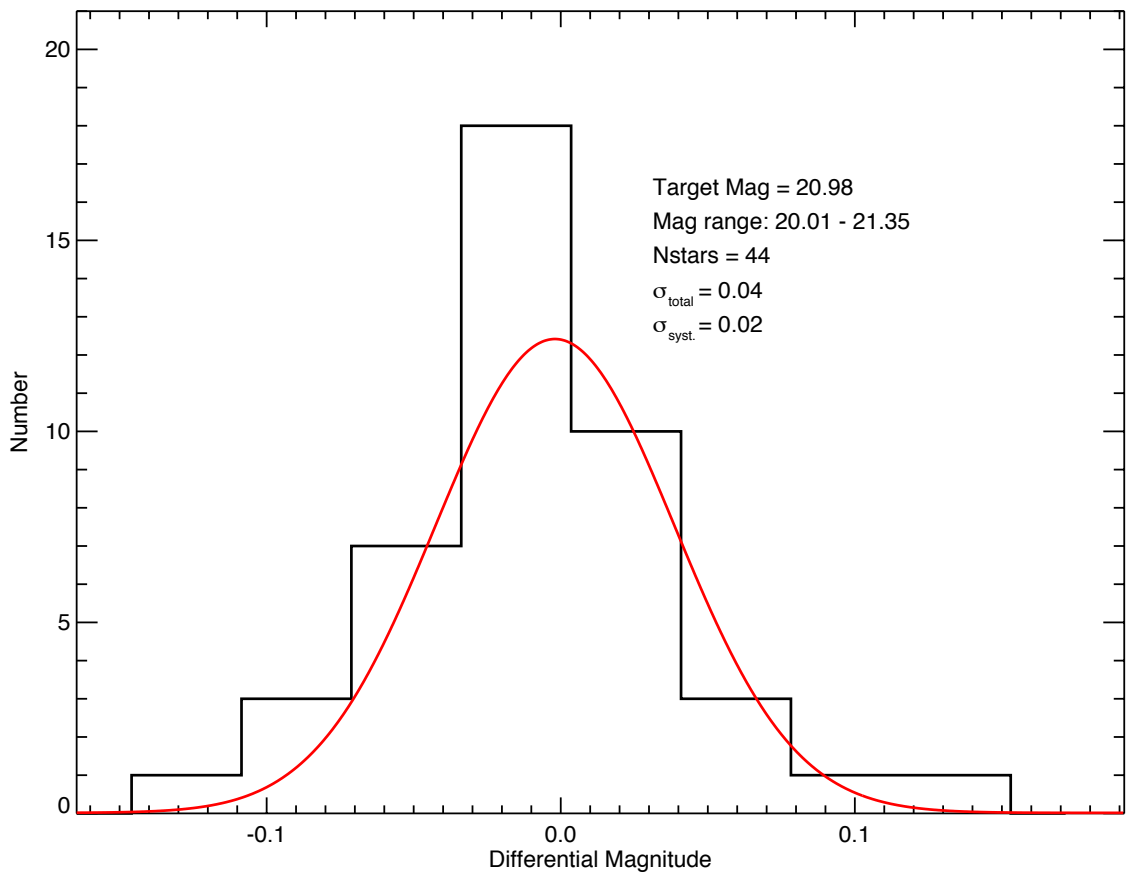}
\caption{Distribution of differential magnitudes relative to the night's median for field stars in Frame 5 comparable 
in magnitude to the target (histogram).  The smooth curve shows a gaussian fit to the distribution.  The gaussian sigma 
($\sigma_{g}$) is given as \sigtot\, in the upper right corner, along with the target's magnitude, range of magnitudes of the comparison stars, number of comparison stars
within this range, and computed systematic error (\sigsys).  The systematic error calculation is described in more detail in Section \ref{sec:processing}.  The differential field star magnitudes are well fit by a gaussian, rendering reasonable systematic error computations.  \label{Fig:syserr}}
\end{center}
\end{figure}

We recognize that systematic errors - a bias in the way the photometry was performed, not in the data itself - may be present due to over- or under-sampling of the background, non-optimal criteria for rejecting contaminants, and/or source flux threshold being too low or high.
To test and correct for systematic errors, after removing the shift between frames for cloud correction, we subtracted each star's median magnitude for the night and fit a gaussian to the distribution of differential star magnitudes in that frame.  The HWHM of the gaussian fit ($\sigma_{g}$) is the same as the total error (\sigtot) and should be a sum in quadrature of the statistical and median systematic error for each target (\sigstat\, and \sigsys, respectively).  Therefore, $\sigma _{sys}^{2} = \sigma_{g}^{2} - \sigma_{stat}^{2}$.  If $\sigma_{g} \lesssim \sigma_{stat}$, then we 
considered $\sigma_{sys} = 0$.  We note that this is the systematic error for field stars, not the untrailed target.  Because the trailed field stars cover a larger number of pixels, the probability of contamination and erroneous background sampling is higher than for the target.  Therefore, these systematic errors may be overestimated for the target.  Figure \ref{Fig:syserr} shows a typical gaussian fit to the differential magnitude distribution.  We performed Monte Carlo simulations of random gaussian distributions with different sample sizes and found that fitting greater than 30 field stars gave a $>95$\% chance of getting a real and reasonable solution.  We found that $\sigma_{g}$ varied depending on the absolute magnitude range of the stars used, with brighter stars giving a smaller $\sigma_{g}$ than fainter stars.  Therefore, to determine the systematic error most representative of the target, we considered only the $\gtrsim 30$ field stars closest in magnitude to the target (R-magnitude $\sim 21.3$).  Lastly, to fit the data to the model, we tested relative offsets in increments of 0.0001 magnitudes within 0.1 magnitudes of the median-subtracted target magnitudes and used the offset that gave the lowest RMS residuals against the model.

\section{Aperture Photometry Algorithms \label{sec:apphotalgorithms}}

Aperture photometry packages usually differ in their definition of aperture size and shape and/or in the way the background is determined, but they are popular largely because they are not sensitive to irregular PSF shapes.  Determining the background correctly is one of the fundamental challenges of photometry.  \cite{1989PASP..101..616H} found that the flux measured within an aperture radius is extremely sensitive to correct background measurement and subtraction.  If the sky is overestimated, then too much background is subtracted, and as the aperture size is increased, the sum of counts within the aperture decreases rather than yielding a constant flux.  This effect is less prevalent for brighter objects, whose PSF wings are more distinct from the noise out to larger radii.  It is therefore recommended not to use a bright star alone to determine which aperture radius captures a large percentage of the source flux ($\gtrsim 95$\%) because a faint object's flux at that radius may already be significantly contaminated by background.  \cite{1989PASP..101..616H} also found that different background measurement methods (\eg median vs. mean, etc.) yield different accuracies.  They found that a weighted mean is more accurate for fainter objects but underestimated uncertainties for brighter objects.

In addition to proper background subtraction, variation in the PSF across the frame caused by optical aberrations is an important consideration with aperture photometry.  Most aperture photometry algorithms, save SExtractor, do not allow for variation in aperture size across a chip, leaving it difficult to choose the best constant aperture radius for the entire frame.  In a detailed study using the HST WFPC-2 CCD camera, \cite{1995astro.ph..3083T} found that radial variation in the  images was significant ($\sim 3$\%), and \cite{2004AN....325..583I} found up to a 15\% difference in FWHM across the CCD cameras for the Sloan Digital Sky Survey (SDSS).  Although these are extreme examples of wide field imagers with large distortions, it is important to check for spatial variations.  We saw no such distortion in our images to a fraction of a percent, so no spatially-variant aperture was necessary.    

Popular aperture photometry algorithms that were compared include the PHOT task within the IRAF/APPHOT package and Source Extractor (SExtractor).  We also implement the tphot aperture photometry technique, optimized in centroiding and reducing background contamination.  Ours is the first publication of the tphot software.  Lastly, the Aperture Photometry Tool (APT) was developed as an educational tool, and because of its ease of use, possible similarity in quality to SExtractor, and application to a wider populace, we include it in our study \citep{2012PASP..124..764L}.  

\subsection{The IRAF/APPHOT PHOT task \label{sec:phot}}

The PHOT task is widely used for traditional aperture photometry \citep{1986SPIE..627..733T}.   An identical task can be found within the DAOPHOT package.  Depending on the sky in the vicinity of the target, there are several opportunities to change the way the background is measured within PHOT, four of which are tested here (each on all 10 images) with IRAF version 2.14.1.  PHOT gives the option of rejecting sky annulus pixels that are above or below a user-defined sigma-threshold, which we chose to be 2.5.   Photometry for both the target and the field stars were done with these settings.

To determine the aperture radius, we measured a curve-of-growth on several stars of different magnitudes spread over the chip and identified the aperture radius that contained 99.5\% of the light, with weighting given to brighter stars.  In general, we found that this aperture only varied by $\lesssim4$\% between bright unsaturated stars and fainter stars at S/N $\sim 30$, indicating that the background was measured accurately.  We repeated this process for all 10 images used and took the largest frame-specific aperture (18 pixels, or $\sim3.9\arcsec$) as the fixed, unweighted aperture radius for all images.  The aperture chosen was used for both the trailed stars and the target.

One concern in choosing a photometric aperture radius for moving targets is that the target PSF should be narrower than trailed field stars.  Consequently, the aperture chosen from field star curves of growth may be larger than the target's 99.5\% flux radius, causing an increased contribution from sky noise and potential faint background source contamination.  

\subsubsection{Centered sky annulus \label{sec:photcentered}}

If the target moves along a track free of significant background contaminants within the sky annulus (see Fig. \ref{fig:images}), then a target-centered sky annulus best represents the sky in the photometric aperture.  For fainter point sources, the PSF wings are indistinguishable from random noise beyond the aperture containing 99.5\% of the target flux (assuming the CCD is flat), so the choice of inner radius of the sky annulus is somewhat arbitrary, so long as it is outside the aperture.  For bright point sources, however, a small amount of noticeable source flux may be in the region outside the aperture, affecting the magnitude uncertainty.  To be sure we were clear of the PSF wings of these bright stars, we defined the inner radius of the annulus at $6\arcsec$, or 28 pixels (10 pixels greater than the aperture).  We chose the outer bounds of the sky annulus to be $11\arcsec$ (50 pixels) to ensure good background sampling.  

\subsubsection{Off-center, manual sky measurement\label{sec:photoffcenter}}

If the sky annulus centered on the target contains a significant amount of unavoidable contamination from background sources, one can manually measure the sky background away from the target.  A manually-defined sky annulus would also be necessary for active comets to avoid dust trails.  We used a sky aperture of the same area as the centered sky annulus in Section \ref{sec:photcentered} and measured the sky 10-15 times in a radially-isotropic distribution within $\sim 17.5\arcsec$ (80 pixels) of the target for every frame, avoiding overlap with the photometric aperture or contaminants.  The final sky and sky noise levels used in the photometry for each frame were the average of the individual measurements in that frame.  Manually measuring the background at several locations off-center from the photometric aperture is too time-intensive to execute for every field star in every frame.  Thus, for field stars only, we used PHOT's non-interactive centered sky annulus technique described in Section \ref{sec:photcentered}.  Since the systematic error is empirically determined from the dispersion in differential field star magnitudes, the systematic error per frame was the same as determined in the centered sky annulus method.  

\subsubsection{Small aperture $+$ aperture correction\label{sec:photapcor}}

Decreasing the photometric aperture excludes some sky noise and nearby contamination but may also exclude source flux.  We used bright field stars with well-defined PSF wings to estimate the aperture correction, or amount of source flux lost by using a smaller aperture.  We then added back the residual magnitude estimated from our aperture correction to the magnitudes measured with the smaller aperture for the target.  With moving targets, we risk that the untrailed target may not be accurately represented by the aperture correction given by trailed field stars, which have a wider PSF than the target.

The choice of the small aperture size is somewhat arbitrary, so long as it is consistent from frame to frame, so we chose a small aperture radius of $\sim 2.2\arcsec$ (10 pixels) to avoid the cosmic ray hit seen in Frame 1 (Fig. \ref{fig:images}).  We measured the aperture correction between the small aperture and a large aperture at $\sim 3.9\arcsec$ (18 pixels), using only field stars with \sigtot$ \leq 0.02$, to ensure good flux sampling in the PSF wings.  Field stars that gave aperture corrections more than 2 standard deviations from the mean for that frame were excluded from the computation, leaving on average 16 field stars per frame to determine the aperture correction.  The standard deviation of these stars' aperture corrections ($\sigma_{ap}$) was used as the error on the frame's aperture correction, which was added in quadrature to \sigstat\, and \sigsys\, to compute \sigtot\, (\ie \sigtot$^2 = $\,\sigstat$^2 + $\,\sigsys$^2 + \sigma_{ap}^{2}$).  To measure the sky, we used the sky annulus settings described in Section \ref{sec:photcentered}.

\subsubsection{Small aperture $+$ aperture correction photometry on \\sky-subtracted images made with IMARITH\label{sec:photIMARITH}}

There are sometimes faint background contaminants close to the target that render inaccurate sky determination or add to the target flux.  These background sources may not be obvious in individual frames, but can be seen in a stacked image.  To correct for these contaminants, we constructed a median-combined sky composite (to remove the moving target), scaled the sky composite to match the mode of an individual frame's background, and subtracted the scaled, aligned composite from each frame using the IRAF ``IMARITH'' task.  The sky composite was made from 8 of the 10 science frames, with frames 8 and 9 excluded because the target's location did not change enough relative to other frames to remove all target flux from the median sky composite.  The subtracted images should be free of faint sources, but residuals from subtraction close to the centroid of brighter sources are often apparent, caused by inaccurate alignment before subtraction and/or the changes in seeing/image quality or non-linear regime of the detector.  We also found that inaccurate alignment before subtraction led to substantial inaccuracies in background measurement, so we used dozens of bright field stars to compute a shift as accurately as possible.  Figure \ref{fig:skysub}b shows one of the individual frames used in our analysis after IMARITH sky subtraction.

Sky subtraction takes place {\it before} performing photometry, so to assess the improvement to the results, we apply the same photometry technique used on frames without sky subtraction and compare the outcomes.  We chose to implement the aperture correction photometry method (Section \ref{sec:photapcor}) because it is believed to produce more accurate results than the centered and off-center sky annulus methods described in Sections \ref{sec:photcentered} and \ref{sec:photoffcenter}.  Field star photometry was performed on the frames before sky subtraction, and photometry on the target as performed after the subtraction.

\subsubsection{Small aperture $+$ aperture correction photometry on \\sky-subtracted images made with ISIS\label{sec:photISIS}}

Section \ref{sec:photIMARITH} describes how sky subtraction can improve photometry by removing background contaminants, but constructing a sky composite is a sensitive function of determining the correct sub-pixel shift between stars in different frames and scaling the composite to match each frame.  The ISIS software package \citep{2000ASPC..203...50A}, which internally determines shifts, median-combines frames, and subtracts the composite from individual images, was designed to do the best possible job of sky composite construction and subtraction.  The main differences between using IMARITH and using ISIS to build a sky composite are that: (1) the ISIS composite must be built from frames with the best seeing, and (2) before subtracting the composite, ISIS convolves it with a kernel solution for the individual frame, matching the FWHM of the sources in the composite to the same sources in the individual frame.  We made the ISIS sky composite from the five frames with the best seeing, with a mean seeing of $\sim 0.99\arcsec$ compared to the remaining five frames' mean seeing of $\sim 1.08\arcsec$.

ISIS can also perform differential photometry after sky subtraction.  Although \cite{2006MNRAS.370..954I} found that using ISIS instead of traditional aperture photometry did not improve photometry results for star clusters, \cite{2000ASPC..203...50A} noted an improvement by up to a factor of 20 in ISIS photometry over DoPHOT.  However, we were not able to fully adapt the ISIS software for moving objects, even when shifted stamps showing the object at the same position were manually constructed and fed into the algorithm.  Therefore, we could not make use of ISIS's photometry capabilities for our moving target; instead, we only used it for sky subtraction.  Figure \ref{fig:skysub} shows that ISIS offers considerable improvement over IMARITH in removing background sources.  The standard deviation within an aperture radius of $\sim4 \arcsec$ (or 18 pixels) at the location of subtracted field stars was reduced by a factor of $\sim 1.7$ by using ISIS rather than IMARITH to perform sky subtraction.  

\subsection{Source Extractor \label{sec:sextractor}}

Source Extractor (hereafter SExtractor) is a photometry algorithm that was designed to quickly identify and classify field objects and to perform aperture photometry on both extended and point sources.  We used SExtractor 2.5.0 \citep{1996A&AS..117..393B}.    

One of the main differences between SExtractor and other aperture photometry algorithms is that SExtractor constructs a background map for the entire image rather than measure the background within a defined annulus for each object.  The background map is locally determined over a defined mesh size, and pixels above a 3$\sigma$ threshold within the mesh are iteratively discarded from the background computation until no pixels above this threshold remain.  If the field is considered crowded (\ie the sky noise drops less than 20\% per iteration), then the final background within the mesh is computed in a different way from an uncrowded background \citep{2005astro.ph.12139H}.  The final background map is a bi-cubic-spline interpolation over all regions.  By visual comparison between the background map and an individual frame, we found that a mesh size of $64\times64$ pixels and a smoothing-region size of 3 gave the most accurate background map.

SExtractor also offers flexibility in the aperture shape.  There are five types of apertures available within SExtractor, all of which were tested in this study: MAG\_APER, MAG\_AUTO, MAG\_ISO, MAG\_ISOCOR, and MAG\_BEST.  The MAG\_APER option lets the user define a fixed circular aperture to place over all identified sources in a frame.  We chose an aperture radius at $4\times$HWHM (which theoretically encompasses 99.994\% of the object's flux) for each frame.  Using the MAG\_AUTO aperture choice will internally determine an aperture at the Kron radius, defined as the radius that contains 90\% of a source's flux, for each identified field source \citep{1980ApJS...43..305K}.  The aperture's ellipticity and position angle with respect to a row is then computed via the second order moment on each object.  MAG\_ISO draws an aperture around all adjacent pixels above a defined detection threshold, making it highly flexible for irregularly-shaped PSFs and extended sources.  We chose a recommended detection threshold of 1.5$\sigma$ above the background \citep{2005astro.ph.12139H}.  The MAG\_ISOCOR aperture type is a crude method of determining an aperture correction to the MAG\_ISO magnitudes, assuming the object's PSF is an axisymmetric Gaussian \citep{2005astro.ph.12139H}.  Lastly, MAG\_BEST is a combination of MAG\_AUTO and MAG\_ISOCOR; when the neighboring source contamination reaches 10\%, MAG\_ISOCOR measurements are returned, otherwise MAG\_AUTO measurements are returned.  

\subsection{The tphot aperture photometry technique\label{sec:tphotap}} 

The tphot routine has three basic components: (1) triggering on a
potential object, (2) aperture photometry, and (3) PSF fitting
photometry.  The trigger function identifies all
pixels that are a local maximum out to a specified radius (nominally 5
pixels) and that exceed a threshold level (specified as an absolute
level or as a S/N above sky level and noise).

Tphot performs photometry on all triggered objects, but this these calculations may be rejected according to a number of tests, including inadequate
S/N, a fitted centroid that moves from the brightest pixel, a profile that is
too broad or narrow, etc.  Tphot offers the option of using an
external file of object positions.

The uncertainty in the photometry is based on a noise model that
is implemented using two parameters, a gain (e$^{-}$/ADU) and a bias level.
The ``bias'' level is subtracted from the image and the noise is then
taken to be the square root of the number of electrons.  The ``bias''
level can be negative if the image preprocessing has already subtracted
a sky level, and it can be augmented according to read noise variance.

The aperture photometry function of tphot performs a fit for the sky
level near an object, and then calculates the total flux within a
specified aperture, less the sky contribution.  For maximum accuracy the aperture should be 1.5 FWHM in radius
\citep{2011PASP..123...58T}.  For robustness in the
presence of neighboring objects, tphot starts by calculating the
median flux in annular rings around the object of interest.  The sky
level is determined by fitting the outermost rings beyond a specified
radius with a constant plus $r^{-3}$ object profile.  The object flux
is calculated as the sum of pixels within the aperture radius, less
the sky level.  This algorithm has proven to be very accurate and
robust for bright objects and fields that may be somewhat crowded.

\subsection{APT\label{sec:apt}}

The Aperture Photometry Tool (version 2.1.9, hereafter APT) is a relatively new algorithm intended as both a professional and an educational tool capable of performing straight-forward and accurate photometry \citep{2012PASP..124..737L}.  Its main feature is its graphical user interface, encouraging visualization of the process and visual checks on aperture and sky annulus choices.  Because of its simplicity, APT was not designed for or tested on crowded fields.  The user defines the sky annulus and the radius, ellipticity, and position angle of the photometric aperture.  There are also options to remove the median or mean background, perform photometry in batch mode, define an aperture correction, and include/exclude manually-selected pixels from the aperture radius.  APT only accepts integer input pixels, then upon the user's request, it internally recomputes the centroid to 0.01 pixel resolution.  The APT help manual states that this sub-pixel centroiding may perform poorly, especially in more crowded fields.  The user can also define rejection limits (in counts) above and/or below which sky annulus pixels are excluded from background determination.

Performing photometry manually for every field source was impractical because we had over 100 objects in our target list, so we chose to operate in batch mode, which meant using the same aperture and sky annulus settings across the entire frame.  We chose a fixed circular photometric aperture of 18 pixels, inner-sky radius at 28 pixels, and outer-sky radius at 50 pixels (the same settings used in Section \ref{sec:photcentered}).  We also chose to perform median background subtraction and make use of the pixel exclusion feature to remove the cosmic ray hit in Frame 1 (Fig. \ref{fig:images}).  We used the APT-computed sky level and sky sigma in the target's sky annulus to determine the $2.5\sigma$ rejection threshold for background determination in each frame.

\section{PSF-fitting Algorithms\label{sec:psfalgorithms}}

PSF-fitting algorithms differ amongst themselves primarily in the way the model
is stored for later use \citep{1992ASPC...25..297S}.  Some fit an analytic function (Gaussian, 
Lorentzian, or a combination thereof) while others determine an empirical PSF model
through numerical interpolation.  Some PSF-fitting algorithms also offer a spatially-variable solution in case the PSF varies across the chip.  Because PSF-fitting routines theoretically exclude more background contamination from the source flux than an aperture, they offer a potential advantage in accuracy and precision of faint objects \citep{2003BaltA..12..243H}.  

A few comparison studies of PSF-fitting algorithms have been published.  \cite{2007PASP..119.1462B} found that DAOPHOT outperforms DoPHOT by a factor of 1.5-4 in accuracy for fixed targets, but \cite{1991AJ....101.1338F} and \cite{1993PASP..105.1342S} noted that DAOPHOT can be susceptible to background source contamination.  However, the usefulness of these algorithms for moving target photometry has not been explored.  The fact that trailed field stars are used as templates for the PSF model presents a problem to moving targets in that the model will not exactly represent the untrailed target.  We investigate the accuracy and precision of the popular PSF-fitting algorithms DAOPHOT and DoPHOT and the profile-fitting capabilities of the new tphot algorithm.  

\subsection{DAOPHOT \label{sec:daophot}}

DAOPHOT is one of the most popular photometry algorithms.  This software package operates within IRAF and exploits the strengths of both analytic and empirical approaches to PSF-fitting.  First, an analytic function is fit to manually-inspected template stars without overlapping PSFs, and errant pixels in the source PSF (either cosmic ray hits or bad pixels) are down-weighted and local background is determined.  Six different types of analytic functions can be fit - Gaussian, Lorentzian, 2 modified Lorentzians, and 2 Gaussian $+$ Lorentzians \citep{1992ASPC...25..297S}.  The model that returns the lowest scatter is subtracted from each template star, and the average residuals are stored in a two-dimensional lookup table to account for any flux lost in the analytic model.  The user can specify whether or not the PSF varied across the chip, in which case DAOPHOT will compute additional lookup tables representing the linear or quadratic change in flux across the frame.  To ensure that flux is not artificially enhanced or decreased due to spatially-variant lookup tables, the net volume of the higher-order tables is forced to be zero.  The option for a spatially-variable model solution have sometimes been implemented with difficulty, even failing altogether as they did for us \citep{2006MNRAS.370..954I}.  For each field source, the model is scaled down to the flux within a defined fit radius (usually the FWHM), and the PSF model is iteratively fit to field sources.  

We altered the many input free parameters as recommended by the DAOPHOT reference guide, using the DAOEDIT task to determine and define each image's FWHM, background, and sky noise \citep{davis1994reference}.  We defined the inner-sky radius as $4\times$FWHM, the outer-sky radius as $8\times$FWHM, and the radius within which the model computed for template stars is computed as $(4\times$FWHM$)+1$, following the \cite{davis1994reference} suggestions.  Because the target was moving and we were comparing photometry for individual frames, we were unable to use DAOPHOT's ALLFRAME task, which performs photometry on a stacked image \citep{2007PASP..119.1462B,2000PASP..112..177F}.   

\subsection{DoPHOT \label{sec:dophot}}

DoPHOT \citep{1989ESOC...31...69M,1993PASP..105.1342S} was developed as an alternative fitting algorithm to DAOPHOT, because determining the appropriate settings for each image and computing the model PSF for DAOPHOT is very time-consuming.  DoPHOT requires minimal user interaction and optimizes speed; we used version 4.1.
The user inputs estimates for the background, seeing, gain and read noise, which are used by the algorithm to identify sources above a defined threshold.  These sources are fit with an analytical power-law function to determine a best-fit model, which is then subtracted from the frame.  Noise is added back to the black sky patches left after star subtraction to ensure that the patches resemble the sky noise.  The detection threshold is then lowered to fit a model to fainter sources, which are then subtracted.  The brighter sources already identified are added back into the frame and refit, now without possible contamination from faint sources.  The improved model parameters are saved, and the magnitude, background, and poisson noise close to the source is recorded.  These iterations continue for lower and lower thresholds (\ie fainter objects) until no additional stars are detected.  A final model weighted by S/N is computed from all objects that are unambiguously identified as point sources, then all of the objects detected in the image are refit with the improved model.  
We used DAOEDIT's background and FWHM measurements (see Section \ref{sec:daophot}) for the DoPHOT input background and seeing estimates, respectively.  Otherwise, we preserved the default setting as recommended by the DoPHOT manual \citep{1989ESOC...31...69M}.  

\subsection{The tphot PSF-fitting technique\label{sec:tphotpsf}} 

After performing an aperture photometry calculation (Section \ref{sec:tphotap}), tphot also
performs a PSF fit to each potential object.  Using the position and
crude FWHM from the aperture fit, tphot does a least-squares fit of
an elliptical ``Waussian'' (Gaussian truncated at the $r^6$ term)
profile.  Tphot can be requested to perform profile fits of two
parameters (flux and sky), four parameters (plus position), or seven
parameters (plus PSF shape).  It also can fit a trailed profile,
substituting trail length for major axis.  The resulting major axis
$a$, minor axis $b$, and peak value $P$ can be multiplied to obtain
a value $a\,b\,P$ that is proportional to the total flux of the
Waussian.

Depending on the application, tphot may be run multiple times.
It is most commonly run once at a relatively high S/N threshold
to obtain aperture photometry and PSF profile shape parameters
for bright stars.  It is then rerun at a low S/N threshold and
a four parameter fit, forcing all stars to a common PSF profile.
The $a\,b\,P$ value provides accurate, relative photometry for
faint objects as well as bright ones, and the aperture photometry
of bright stars from the previous iteration provides net fluxes.

In this image there are a comparable number of galaxies and stars; a
  different galactic latitude or magnitude limit would be different.
   We did not attempt any automatic star-galaxy classification for this
   application.  Instead, we chose the objects whose major axis a falls
   below the median, and then use the median values of those objects for $a$, $b$, and $\theta$ as representative of stellar PSFs.  A mode of all values less than the median $a$, $b$, and $\theta$ would have also been suitable.  Measuring the change in aperture photometry of stars between different
frames then correcting for the change put all images on the same photometric scale.  The RMS
for these comparisons was less than 0.01 magnitude.

Although the target was trailed, its trailing amount was small enough relative
to the seeing that we did not bother to use the trailed profile
function of tphot.

\section{Photometry Algorithms not Explored here\label{sec:notexplored}}

We were not able to test all existing photometry algorithms.  \cite{2007PASP..119.1462B} find that the Photo software used on the Sloan Digital Sky Survey performs high-quality photometry on both extended sources and field stars, but it is not portable or flexible and was thus not available to install.  Three other prominent photometry packages, ISIS, MOMF, and DIA, could not be modified for use on data for a moving target, though they did offer promising improvements to our photometry results \citep{1992PASP..104..413K,2000ASPC..203...50A,2000AcA....50..421W,2010JASS...27..289L}.  These packages all perform photometry on fixed objects in median-sky subtracted images, identifying low-level deviations from the nightly median.  The main advantage of these methods over traditional sky-subtraction is the process of convolving a ``best-seeing'' sky composite with the FWHM of individual frames before subtraction, so that the PSFs match and no residual background source flux remains on the chip after subtraction, with the exception of saturated pixels (see Fig. \ref{fig:skysub} and Section \ref{sec:photISIS}).  Because the change in magnitude is measured from sky-subtracted images, no absolute magnitude calibration is offered in any of these methods.  These difference-image techniques may also present a challenge to moving targets in that the FWHM would not be the same for the target as the trailed stars, so the convolution would be incorrect and residual flux would always be present.  Section \ref{sec:photISIS} details how we were unable to adapt the promising ISIS software for our moving target.  We assumed we would experience the same difficulties implementing the other difference image techniques, so we did not attempt to run them.  The potential of these difference-frame algorithms to optimize accuracy and precision begs that they be made adaptable to moving objects.

\section{Results\label{sec:discussion}}

For each algorithm, we use the RMS of the residuals against the light curve model as a diagnostic for magnitude accuracy, and the \chisqr\, statistic as a measure of how reasonable the magnitude uncertainties are.  Section 2 explains the model's origin.  Figure \ref{fig:summary} plots the results of all algorithms against the model light curve, shown in order of least to greatest RMS of the residuals.  We also report  \sigsys\, to determine how much systematic bias is incurred in each algorithm and \sigtot\, (calculated using the \sigsys\, determined for each frame and the target-specific \sigstat\, reported by each algorithm), which assesses the precision of the results.  If the magnitudes themselves are accurate and all systematic errors have been accounted for, then \chisqr$\;\sim 1.0$.  If \chisqr$\;\lesssim 1.0$, then the errors (including systematics) are overestimated, but if \chisqr$\;\gtrsim 1.0$, either the magnitudes are inaccurate or \sigtot\, is still underestimated.  Table \ref{tab:stats} lists the RMS of the residuals, the \chisqr, \sigtot, and \sigsys\, for each implementation.

\subsection{PHOT algorithms\label{sec:discphot}}

Figure \ref{fig:summary} and Table \ref{tab:stats} show that the widely-used aperture photometry algorithm PHOT generally performed poorest of all other algorithms tested, with or without subtracting a sky composite with IMARITH.  We saw no change in the results by using a manually-placed, off-center sky aperture as opposed to a centered sky annulus, presumably as long as the sky measurements were made relatively close to the target (Fig. \ref{fig:summary}m vs. Fig. \ref{fig:summary}n).  Aperture correction gave an improvement in RMS by a factor of 2.5 over using an aperture that encompassed 99.5\% of the flux, making it comparable to SExtractor results.  Since the sky was measured in PHOT in the same way before or after sky subtraction, the improvement in RMS must be caused by a more accurate accounting of source flux.

The two PHOT algorithms that involved subtracting a sky composite to remove faint background contamination before performing photometry (Sections \ref{sec:photcentered} and \ref{sec:photoffcenter}) generally rendered a larger light curve amplitude than the model, which increased the RMS of the residuals and \chisqr\, (Fig. \ref{fig:summary}f and Fig. \ref{fig:summary}o).  Visual inspection shows that the ISIS-processed images contain one exceptionally deviant data point at $\phi = 0.332$.  This data point corresponds to Frame 1 in Figure \ref{fig:images}, where a cosmic ray hit that was missed in initial cosmic ray correction is seen close to the target.  It may be that the cosmic ray hit compromised the photometry more so in the ISIS-treated frames than in results from other realizations of PHOT, though there is no clear explanation for why this would be the case.  Also, the cosmic ray removal routine that we used obviously did not correct for all hits because it was designed to flag point source cosmic ray hits (\eg single pixels) rather than grazing hits such as the one in Fig. \ref{fig:images}.  Thus, the routine may have left additional faint contaminants near the target PSF and affected the accuracy of the photometry.
Only after removing the anomalous data point at $\phi = 0.332$ from analysis did ISIS sky-subtraction render a RMS comparable to the other PHOT aperture correction algorithms.  Subtracting the sky with IMARITH before applying an aperture correction offered no improvement in RMS, likely because contaminants were already sufficiently removed from source flux measurements by using a smaller aperture.  We speculate that if the background within the sky annulus contained more faint sources, then performing IMARITH sky subtraction would offer further improvement in accuracy before using aperture correction.   

The \chisqr\, $\gtrsim 1.0$ for all PHOT algorithms, especially the two algorithms involving sky subtraction (Table \ref{tab:stats}), comes from incorrect magnitude measurement, background determination, and/or noise models.  To test whether or not the background was correctly determined and subtracted, we repeated the centered sky annulus method (Section \ref{sec:photcentered}) with successively larger apertures; if the background was overestimated, then the magnitude measured would become steadily fainter toward larger radii, and vice versa \citep{1989PASP..101..616H}.  We found that the magnitude stayed the same, only the magnitude uncertainty changed using larger aperture radii, implying that background subtraction was done correctly.  

PHOT calculates the magnitude uncertainty from the photon noise within the aperture and the standard deviation within the sky annulus, and it includes a term to account for uncertainty in the background level.  Since we find that background subtraction was done properly, we assume that the standard deviation measurement within the sky annulus is also correct.  With sky subtraction, we found that the standard deviation of the background was reduced by a factor of 2, which consequently gave a smaller magnitude uncertainty and even larger \chisqr.  If the aperture contained contamination from background sources, however, then both the source flux and aperture's photon noise would be inaccurate, the magnitude more so.  Such an effect would manifest as slightly large \sigtot\, and even larger RMS, which is what we see in the PHOT results.  The factor of $\sim 2.5$ improvement in RMS by using a smaller photometric aperture but measuring the background in the same way provides further evidence that background contamination within the aperture is significant.  

\subsection{APT\label{sec:discapt}}

\cite{2012PASP..124..764L} compared APT to SExtractor's MAG\_APER, noting overall agreement between the two algorithms.  We do not find the same result here, possibly indicative of APT's limitations toward fainter objects.  Table \ref{tab:stats} shows that APT's residual RMS is roughly twice that of MAG\_APER, though the \chisqr\, is slightly larger due to the slightly smaller \sigtot\, and much larger RMS.  To investigate the cause of the large RMS, we tested for correct background determination by checking that the source flux stayed constant in several aperture radii beyond the original 18-pixel radius.  We saw that as we increased the aperture size, the source flux continued to decrease, indicating that the background was overestimated.  Incorrect background determination can lead to inaccurate measurement of source flux and background uncertainty, which then increase the RMS and \chisqr.  Some photometry inaccuracy may stem from incorrect sub-pixel shifts, as noted in the APT manual \citep{2012PASP..124..737L}.  Like PHOT, APT also appears to do a poor job of systematically excluding faint background contamination from the aperture.  We found that APT measured the largest target flux within a 10-pixel radius of all aperture photometry algorithms, even with the background overestimated, implying background contamination.      

\subsection{DoPHOT\label{sec:discdophot}}

The DoPHOT results are given in Figure \ref{fig:summary}l and Table \ref{tab:stats}.  The comparably large RMS shows that DoPHOT did not do a good job of reproducing the model, performing worst of the PSF-fitting algorithms but slightly better than PHOT's non-aperture correction algorithms.  Magnitude uncertainties are also relatively large compared to the algorithms tested here. We suspect that the main sources of error for both magnitudes and uncertainties are that (1) DoPHOT PSF-fitting fails at accurately modeling the PSF of a trailed field star, and (2) DoPHOT is too inclusive in PSF model computation.  If a field contains many irregularly-shaped, compromised, and/or extended sources, then the final model may be significantly affected.   Due to DoPHOT's limited user interaction, we could not further explore the role of potentially incorrect background determination, insufficient uncertainty estimates, and/or contamination in increasing the RMS and \chisqr.  The \sigsys\, was also the largest of all algorithms tested, indicating that a substantial amount of systematic bias was incurred.

\subsection{SExtractor\label{sec:discsextractor}}

Table \ref{tab:stats} shows that SExtractor improved the RMS by a factor of $\sim2.5$ over PHOT (except for the aperture correction algorithm) and $\sim 2.0$ over APT and DoPHOT, implying better magnitude accuracy.   Of the SExtractor aperture choices, MAG\_ISOCOR delivered the lowest RMS but relatively unchanged \sigtot, meaning it offered an improvement in accuracy but not in precision.  MAG\_ISOCOR determines an aperture correction to the MAG\_ISO magnitude measurements (Fig. \ref{fig:summary}h), and the aperture correction technique was already shown in Section \ref{sec:discphot} to be more accurate than using a large aperture to add up source photons, so it is unsurprising that MAG\_ISOCOR provided an improvement in RMS and \chisqr\, over MAG\_ISO (Table \ref{tab:stats}).  \cite{2007PASP..119.1462B} noted that in isophotal mode, the position and shape of sources detected in SExtractor may be systematically inaccurate toward fainter magnitudes, but given the success of MAG\_ISO and MAG\_ISOCOR here, our target must not have been faint enough to be affected. 

The MAG\_APER choice (Fig. \ref{fig:summary}j) gave the smallest \chisqr, owing to it having a relatively large \sigtot\, and a minimally different RMS.  This finding suggests that a fixed circular aperture can include significant amounts of background, adding more background uncertainty to the final error.  The MAG\_AUTO aperture option (Fig. \ref{fig:summary}g) offered essentially no improvement in accuracy but improvement in precision by a factor of $\sim 2$ over MAG\_APER, implying that it was more effective at excluding pixels that do not contain source flux.  The MAG\_BEST results were almost identical to the MAG\_AUTO results, meaning that the field was relatively uncrowded (Table \ref{tab:stats}).

Despite SExtractor's general success over other aperture photometry algorithms (except tphot) at accurately reproducing the model light curve, the \chisqr\, was comparably large.  A smaller RMS but larger \chisqr\, suggests that the magnitude uncertainties are underestimated.  \cite{1996A&AS..117..393B} pointed out that the uncertainty in the local background estimate (which is notably complex) was not included in the final reported magnitude.  We therefore assume that this exclusion made the magnitude uncertainties in our experiments artificially and significantly small.  Because the degree to which the magnitude uncertainties are artificially small is unknown, it is difficult to determine whether or not inaccuracy of the SExtractor magnitudes also contribute to the large \chisqr\, value. 

\subsection{DAOPHOT\label{sec:discdaophot}}

The DAOPHOT results are given in Figure \ref{fig:summary}c and Table \ref{tab:stats}, which shows that DAOPHOT is the third-most accurate algorithm tested, behind the two tphot algorithms.  The \sigtot\, is comparably small, indicating good isolation of source flux, though the small \chisqr\, value may indicate that the magnitude uncertainty is still overestimated.  \cite{2007PASP..119.1462B} noted that systematic errors stem from inadequate correction factors in the wings of the model PSF, and that DAOPHOT consistently underestimated the uncertainties by $\sim 20$\%, which is inconsistent with our results.  We suspect that the underestimation they noted was for brighter objects, where the intricacies of the true PSF may be poorly reproduced by the model.  The \sigsys\, in our results was relatively small, suggesting minimal systematic bias compared to other algorithms.  

The somewhat small amount of trailing in the template stars did not seem to affect the accuracy of the PSF model's fit to the target.  We hypothesize that because the trailing was within the expected seeing disc, DAOPHOT's analytic functions had no difficulty computing a model that was applicable to both the trailed template
stars and the untrailed target.  We tried using DAOPHOT to build a model from moving-object frames where the stars
showed more pronounced trailing ($\sim 3.7\arcsec$, or trailing aspect of $\sim 5$), but the \chisqr\, fit of the model to the
untrailed target was $\sim 6-20$ per frame, clearly indicating that the template stars were
too trailed to reasonably match the target's PSF.  In contrast, the average \chisqr\, of DAOPHOT fits to the \toss\, data was 1.84.  

\cite{2007PASP..119.1462B} found that performing aperture photometry within DAOPHOT (the same as aperture photometry with PHOT described in Section \ref{sec:phot}) rendered more accurate results than DAOPHOT's PSF-fitting routine because the flux of faint objects is increasingly underestimated toward fainter magnitudes.  Because our results do not reflect this flux inaccuracy, we assume that \toss\, was bright enough (S/N $\sim 25$ from DAOPHOT) for this not to be a factor.  

\subsection{tphot \label{sec:disctphot}}

Results from the tphot PSF-fitting and aperture photometry algorithms are shown in Figures \ref{fig:summary}a and \ref{fig:summary}b, respectively, and in Table \ref{tab:stats}.  The low RMS value shows that tphot produced the most accurate results of any algorithm tested.   

The tphot PSF-fitting algorithm returned the lowest \sigtot\, of all algorithms tested, implying excellent precision, but \chisqr\, $> 1.0$ suggests that the errors may be very slightly underestimated.  Visual inspection of the results in Fig. \ref{fig:summary}a shows that the uncertainties on magnitudes are consistent with the model, except for one data point at $\phi = 0.587$ (corresponding to Frame 3 in Fig. \ref{fig:images}), which is the likely cause of the large \chisqr.  We can find no reason to exclude this point from analysis unless the model is poorly constrained at this particular rotation phase or the target PSF contains unavoidable deviant pixels.  Removing this point gives \chisqr\, $ = 0.99$, exactly what it should be if the data accurately represent the model, so either the questionable data point is compromised in some way or tphot is reporting inaccurately.  Given the success of tphot at reproducing the model at all other phases, we assume the problem is not with the algorithm itself and rather with the data, possibly due to a bad pixel missed during bad pixel mask construction.  Despite the target PSF not appearing especially radially-isotropic in the images (\eg some frames were out of focus; Fig. \ref{fig:images}), tphot did not exclude asymmetric flux within the circular aperture annuli by computing the median.  

Tphot's aperture photometry gave a \chisqr\, $\sim 1.4$ for all data and \chisqr\, $\sim 1.0$ if we exclude the data point at $\phi = 0.587$, despite a relatively small \sigtot, meaning the magnitudes and uncertainties are a good representation of the model.  We also note that the low \sigsys\, implies that the tphot aperture photometry technique incurs little systematic bias compared to all algorithms explored (except DAOPHOT, which has a similarly low mean \sigsys).  For these reasons, we have no reason to suspect any errors in the tphot aperture photometry routine.

\section{Summary and Discussion\label{sec:summary}}

The tphot algorithm produces the best photometry results for our faint moving target, improving RMS very minimally over DAOPHOT, and by a factor of $\sim 1.9-2.3$ over SExtractor, $\sim 4.3$ over APT, $\sim 4.7$ over DoPHOT, and $\sim 2.0-8.5$ over IRAF's PHOT.  Tphot's success is likely thanks to its careful treatment of the centroid, exclusion of contamination within the photometric aperture, and use of an iteratively-refined Waussian function.  An updated version of tphot is currently under construction to better handle significantly trailed objects.  DAOPHOT produced the next-best results, giving the same accuracy as tphot's aperture photometry algorithm (largely due to careful accounting of source flux) but overestimated errors.  While offering a distinct advantage in speed, SExtractor generally underestimates magnitude uncertainties, likely because of exclusion of error in the background determination.  The MAG\_ISOCOR aperture technique renders the most accurate and precise magnitude measurements amongst the SExtractor options because as an aperture correction technique, it more carefully isolates target flux and excludes contaminants.  IRAF's PHOT algorithm can only accurately measure magnitudes of a moving target if an aperture correction technique is used (without sky subtraction), though it returns median uncertainties $\sim 4$ times larger than what could be achieved with tphot.  Although the background was measured correctly, PHOT did not exclude background contaminants from within the aperture, giving inaccurate photometry.  The APT algorithm, while fairly flexible and excellent as a visualization and educational tool, was outperformed by all other software except DoPHOT and a few PHOT algorithms.  APT struggles with both accurately measuring the background and with systematically excluding contaminants from within the photometric aperture.  We suggest that APT is better suited for photometry on brighter objects.  DoPHOT performed slightly worse in both accuracy and precision compared to APT.  We deduce that its fitting routine was not able to as accurately produce a PSF model possibly because it included too many field sources in its construction of a model PSF and/or it was very sensitive to trailed PSFs.  

At some point, trailing will be significant enough for the PSF peak to be two-dimensional (\ie a line rather than a point).  In these cases, the ideal aperture would be the same shape of the target PSF to exclude as much background contamination as possible.  Techniques that only allow a circular aperture such as tphot, SExtractor's MAG\_APER, and PHOT will incorporate large amounts of background contamination with an aperture that encompasses the extent of the trailed PSF.  The elliptical aperture allowed by APT and SExtractor's MAG\_AUTO and MAG\_BEST may be a good approximation to the target shape if trailing is minimal (\eg our data), but at some point, trailing may be significant enough for an ellipse's axes to extend well beyond the PSF.  The only aperture photometry technique that we tested that places no restrictions on target shape are SExtractor's MAG\_ISOCOR and MAG\_ISO, which we consequently predict will be most suitable of the techniques tested here for aperture photometry of significantly trailed objects.  

These findings could be more statistically robust if we had more images of \toss, which would eliminate the dependency on a relatively small sample size of N $= 10$.  As such, our results are only applicable to data sets similar to ours -- for slow-moving objects with S/N $\sim 35$ (according to uncertainties given by our most precise algorithm), mostly because no other images for a moving object with a well-determined light curve described by a shape model were available to us at the time.  It would be useful to test these algorithms on other objects with (1) brighter targets to determine at what point aperture photometry starts to outperform PSF-fitting because of more well-defined intricacies in the PSF morphology that are not reproduced by analytic functions, and (2) a range of trailing aspects to determine the aspect beyond which photometric accuracy becomes significantly compromised.  \cite{2012PASP..124.1197V} tested the latter for a 2-dimensional symmetric gaussian function and a square aperture 3$\times$FWHM, finding a $\sim 1.0$ and 0.5 magnitude loss, respectively, at a trailing length of 3$\times$FWHM, but this experiment should be repeated for popular algorithms.

PSF-fitting routines suffer similar challenges to accurately representing the target PSF as aperture photometry.  Techniques that fit a radially-isotropic function to the data will fail in recovering the true PSF shape, but no profile-fitting technique that we tested makes this assumption of radial isotropy.  Allowing for an elliptical analytic function is a first-order approximation to representing the true PSF shape of minimally-trailed data, and all of the PSF-fitting techniques we tested have this feature.  However, rather than a single function calculated at the centroid, a trailed PSF might be best represented by the sum of a series of analytic functions calculated at different increments along the direction of motion, such as a trailed Gaussian given in its analytic form by \cite{2012PASP..124.1197V}.  However, this technique has not yet been adapted into software available to the public.  Existing PSF-fitting algorithms cannot calculate a series of different functions for a single trailed source, so we suspect that in the event that an ellipse is no longer a good approximation to the PSF shape and until trail-fitting algorithms are made available, all PSF-fitting packages will fail at accurately representing the PSF.  Therefore, of all the techniques tested here, SExtractor's MAG\_ISO and MAG\_ISOCOR are likely the only ones capable of performing accurate photometry on significantly trailed PSFs.

\acknowledgments

This material is based upon work supported by the National
Aeronautics and Space Administration by NASA Grant Nos.
NAG5-4495, NNX07A044G, NNX07AF79G,
and from the National Science Foundation through grant
AST-1010059.  Partial support for this work was provided by National Science
  Foundation grant AST-1009749.  Initial image processing in this paper has been performed using the
IRAF software. IRAF is distributed by the National Optical Astronomy
Observatories, which is operated by the Association of Universities
for Research in Astronomy, Inc. (AURA) under cooperative
agreement with the National Science Foundation.  %We thank the people of Hawai'i for the use of their sacred space atop Mauna Kea.  

%\bibliography{sonnettbib.bib}

\clearpage

\begin{landscape}
\begin{center}
%\begin{rotate}
\begin{figure}[h]
\includegraphics[angle=0,width=8.3in]{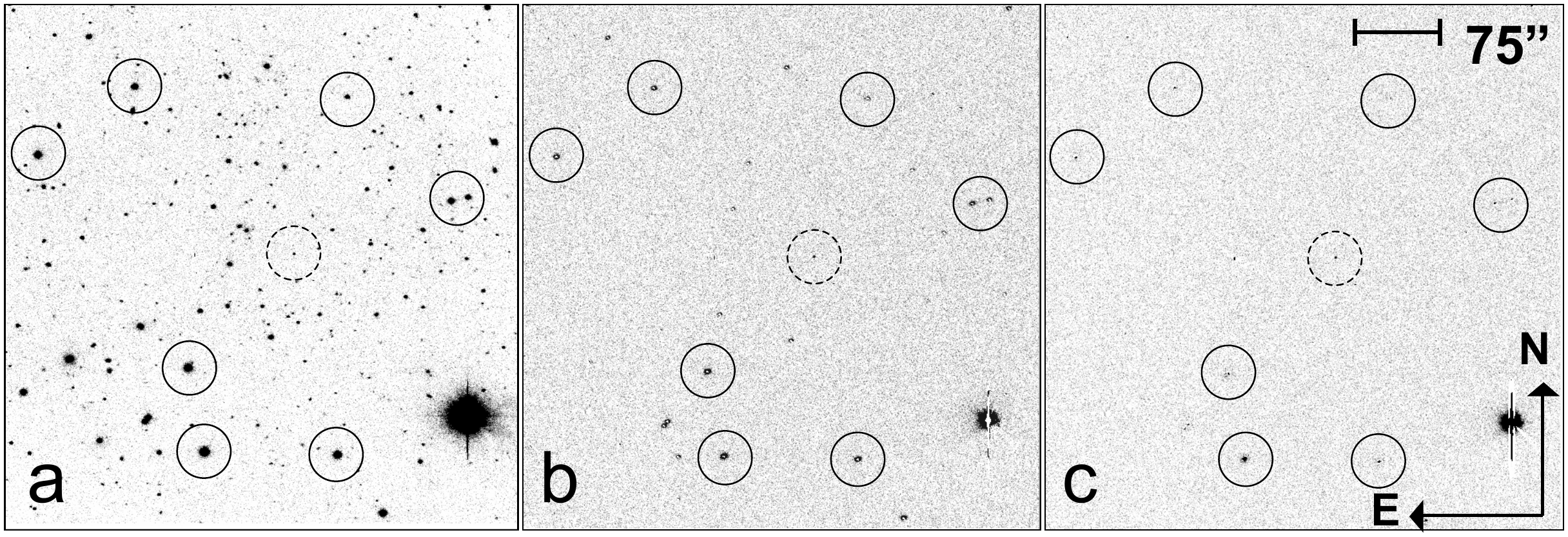}
\caption{A comparison between sky composite subtraction results for a full frame (Frame 7), with a circle placed around several relatively bright field stars visible in the raw images (Sections \ref{sec:photIMARITH} and \ref{sec:photISIS}).  The moving target has a dashed circle placed around it in each panel.  The size and orientation is the same for all panels.  a) The original frame before sky subtraction.  b) The residuals of composite subtraction performed with IRAF's IMARITH task, which subtracted a median-combined reference sky image from individual frames.  The negative spots are residuals left from field stars after subtraction and are likely caused by inadequate alignment and/or composite scaling before subtraction.  c) Sky subtraction residuals using ISIS software, which convolves a ``best-seeing'' sky composite with a spatially-variable kernel representative of the individual frame's FWHM in order to match the image quality before subtraction.  ISIS subtraction shows a noticeable improvement over IMARITH subtraction in fully removing background sources. \label{fig:skysub}}
\end{figure}
%\end{rotate}
\end {center}
%\end{landscape}

%\clearpage

%\begin{landscape}
\begin{center}
%\begin{rotate}
\begin{figure}[h]
\includegraphics[angle=0,height=5.15in]{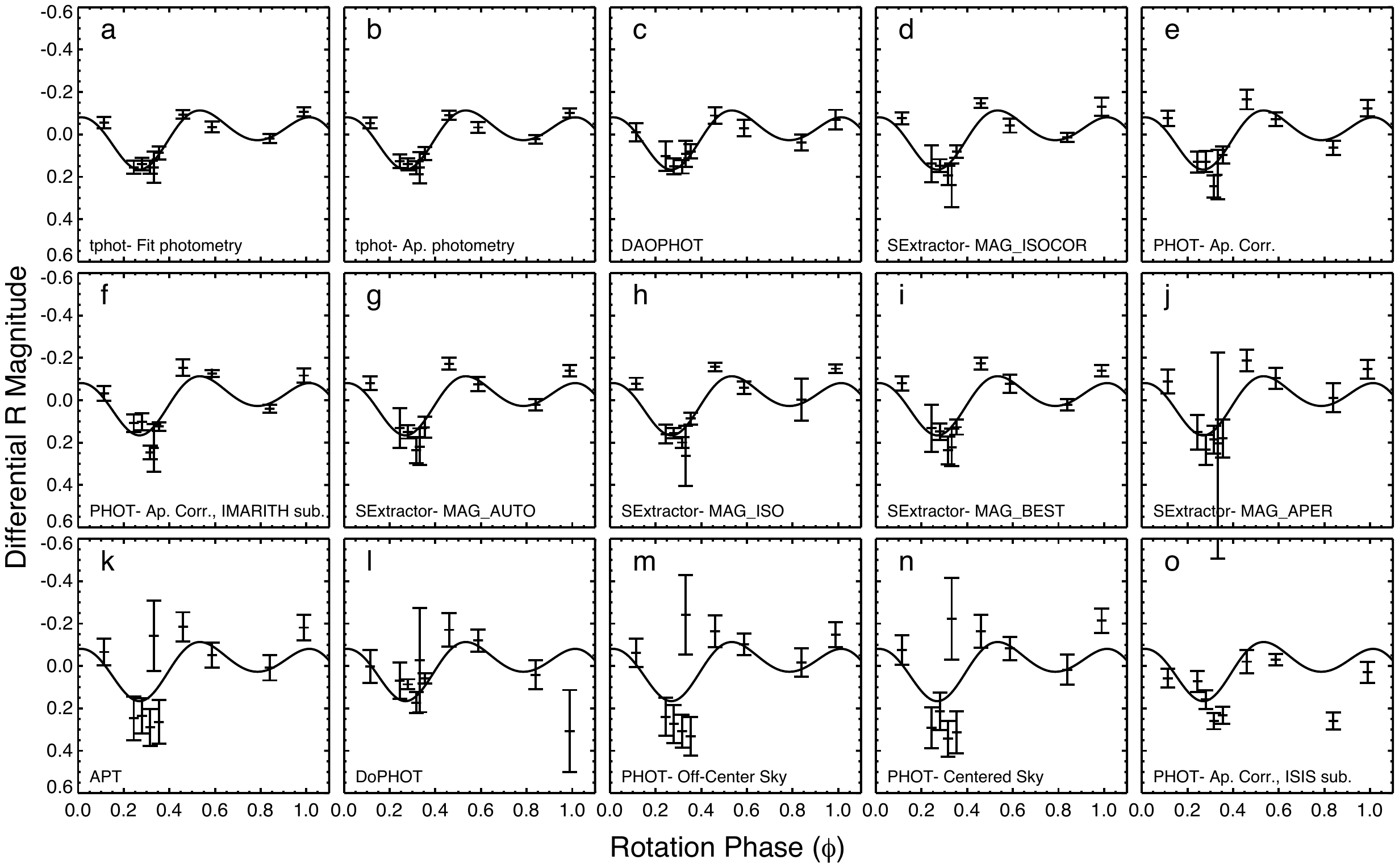}
\caption{Photometry experiment results.  The panels are arranged in left-to-right, top-to-bottom order of smallest to largest RMS of the residuals against the model, which we use as a diagnostic for accuracy.  The algorithm name is given at the bottom of each panel, and pertinent statistics are given in Table \ref{tab:stats}.   \label{fig:summary}}
\end{figure}
%\end{rotate}
\end{center}
\end{landscape}

\begin{center}
\begin{table}
\caption{Statistics for each photometry algorithm.  We give RMS residuals against the model, the reduced chi-squared (\chisqr), the mean total photometric uncertainty (\sigtot), and the mean systematic error calculated (\sigsys).  The results are ordered by increasing RMS. \label{tab:stats}}
\begin{tabular}{llllll}
\tableline\tableline
& Panel & & & Mean & Mean \\
Technique & in Fig. \ref{fig:summary} & RMS\tablenotemark{a} & \chisqr\, & \sigtot\, & \sigsys\,  \\
\tableline
tphot, Fit photometry		    & a & 0.030 & 1.436 & 0.031 & 0.012 \\ 
tphot, Ap. photometry		    & b & 0.031 & 1.488 & 0.032 & 0.008 \\ 
DAOPHOT				    & c & 0.031 & 0.561 & 0.044 & 0.010 \\ 
SExtractor, MAG\_ISOCOR		    & d & 0.058 & 2.501 & 0.045 & 0.027 \\ 
PHOT, Ap. Corr.			    & e & 0.060 & 1.855 & 0.049 & 0.021 \\ 
PHOT, Ap. Corr., IMARITH Subtraction & f & 0.062 & 3.296 & 0.039 & 0.018 \\  
SExtractor, MAG\_AUTO		    & g & 0.063 & 2.985 & 0.047 & 0.023 \\ 
SExtractor, MAG\_ISO		    & h & 0.064 & 5.062 & 0.046 & 0.032 \\ 
SExtractor, MAG\_BEST		    & i & 0.064 & 3.008 & 0.050 & 0.028 \\ 
SExtractor, MAG\_APER		    & j & 0.068 & 1.342 & 0.101 & 0.066 \\ 
APT				    & k & 0.125 & 1.858 & 0.085 & 0.026 \\ 
DoPHOT				    & l & 0.141 & 2.145 & 0.090 & 0.082 \\ 
PHOT, Off-Center Sky Annulus	    & m & 0.159 & 2.413 & 0.085 & 0.048 \\ 
PHOT, Centered Sky Annulus	    & n & 0.161 & 2.655 & 0.089 & 0.048 \\ 
PHOT, Ap. Corr., ISIS subtraction   & o & 0.255 & 14.840 & 0.048 & 0.025 \\ 
\tableline
\end{tabular}
\tablenotetext{a}{The RMS of the residuals against the model.}
\end{table}
\end{center}

\end{document}